\documentclass[%
 reprint,
superscriptaddress,
nofootinbib,
 amsmath,amssymb,
 aps,
pre,
longbibliography
]{revtex4-1}
\newcommand{\tabitem}{~~\llap{\textbullet}~~}
\usepackage{dcolumn}
\usepackage{color}
\usepackage{amsmath,amsfonts,amssymb,amsthm, tabu}
\usepackage{xspace}
\usepackage{hyperref}
\usepackage{dsfont}
\usepackage{tikz}
\usepackage{stmaryrd}
\usepackage{mathrsfs}
\usepackage{lipsum}
\usepackage{braket}

\makeatletter

\newcommand{\avg}[1]{\langle #1 \rangle}
\newcommand{\norm}[1]{\left\vert#1\right\vert}
\newcommand{\abs}[1]{\left\vert#1\right\vert}

\makeatother
\numberwithin{equation}{section}

\begin{document}

\title{By Force of Habit: Self-Trapping in a Dynamical Utility Landscape}

\author{Jos\'{e} Moran}
\email{jose.moran@polytechnique.org}
\affiliation{Centre d'Analyse et de Math\'{e}matique Sociales, EHESS, 54 Boulevard Raspail, 75006 Paris}
\affiliation{Chair of Econophysics and Complex Systems, Ecole polytechnique, 91128 Palaiseau Cedex, France}

\author{Antoine Fosset}%
\affiliation{Chair of Econophysics and Complex Systems, Ecole polytechnique, 91128 Palaiseau Cedex, France}
\affiliation{LadHyX UMR CNRS 7646, Ecole polytechnique, 91128 Palaiseau Cedex, France}

\author{Davide Luzzati}
\affiliation{Chair of Econophysics and Complex Systems, Ecole polytechnique, 91128 Palaiseau Cedex, France}
\affiliation{LadHyX UMR CNRS 7646, Ecole polytechnique, 91128 Palaiseau Cedex, France}

\author{Jean-Philippe Bouchaud}
\affiliation{Capital Fund Management, 23 Rue de l'Universit\'{e}, 75007 Paris\medskip}
\affiliation{Chair of Econophysics and Complex Systems, Ecole polytechnique, 91128 Palaiseau Cedex, France}

\author{Michael Benzaquen}%
\affiliation{Chair of Econophysics and Complex Systems, Ecole polytechnique, 91128 Palaiseau Cedex, France}
\affiliation{LadHyX UMR CNRS 7646, Ecole polytechnique, 91128 Palaiseau Cedex, France}
\affiliation{Capital Fund Management, 23 Rue de l'Universit\'{e}, 75007 Paris\medskip}

\date{\today}
\begin{abstract}
Historically, rational choice theory has focused on the utility maximization principle to describe how individuals make choices. In reality, there is a computational cost related to exploring the universe of available choices and it is often not clear whether we are truly maximizing an underlying utility function. In particular, memory effects and habit formation may dominate over utility maximisation. We propose a stylized model with a history-dependent utility function where the utility associated to each choice is increased when that choice has been made in the past, with a certain decaying memory kernel. We show that self-reinforcing effects can cause the agent to get stuck with a choice by sheer force of habit. We discuss the special nature of the transition between free exploration of the space of choice and self-trapping. We find in particular that the trapping time distribution is precisely a Zipf law at the transition, and that the self-trapped phase exhibits {\it super-aging} behaviour. 
\end{abstract}

\maketitle

\section{Introduction}

A key assumption in rational choice theory is that individuals set their preferences according to an utility maximization principle. Each choice an individual can make is assigned a certain ``utility'', i.e. a quantity measuring the satisfaction it provides to the agent and frequently related to the dispassionate forecast of a related payoff. This framework is often accompanied by the assumption that the agent considers \textit{all} available choices present to her/him, weighs their utilities against one another, and then makes her/his choice taking into account possible constraints, such as a finite budget.

A number of criticisms to this view of human behaviour have emerged, with e.g. Simon~\cite{Simon1955} as a key figure highlighting that individuals may be ``satisfiers'' rather than pure optimisers, in the sense that there is both a computational cost and a cognitive bias related to considering the universe of available choices. Sometimes finding the optimum of the utility function can itself be such a computationally hard problem that even the most powerful computers would not be able to find it in a reasonable amount of time. This led to the idea of bounded rationality as a way to model real agents~\cite{simon1972theories,selten1990bounded,arthur1994inductive,gigerenzer2002bounded}.
More recently, Kahneman~\cite{kahneman2003psychological,kahnemanth} pointed at what he considers to be significant divergences between economics and psychology in their assumptions of human behaviour, with a special emphasis on the empirical evidence of the cognitive biases, and therefore the {irrationality}, that guides individual behaviour. A pervasive effect, for example, is that the utility of a certain choice strongly depends on the choice made by others. These so called ``externalities'' can lead to interesting collective effects, where choices made by agents synchronise and condense on a small subset of choices, or lead to confidence crises  -- see for example \cite{Durlauf,durlauf2004social, Borghesi2007, bouchaudcrisis2013, morelli2019}.  

An interesting idea developed in~\cite{Kahneman2000} is the fact that the utility associated to a certain decision may depend also on our memory if it has already been made in the past. Here we propose a simple model that encapsulates this idea, and show that this too can lead to choices that do not necessarily conform to their ``objective'' utilities, but are rather dominated by past choices alone. This is related to what economists call ``habit formation'' \cite{campbell1999habit,abel1990asset,constantinides1990habit,carroll2000saving, fuhrer2000habit,pol1970sav}. Memory effects chisel the utility landscape in a way that may render objectively sub-optimal choices subjectively optimal. In the case of sufficiently long range memory, agents may, in a self-fulfilling kind of way, become ``trapped'' forever in a certain choice and stop exploring alternative choices.  

As a practical example, one may imagine a situation where one must choose where to have lunch every day. Standard rational theory dictates that we ought to scrutinize every restaurant, eatery and cafeteria, taking into account our personal tastes and the costs associated with going to any of these places. In contrast, we want to model the fact that habit may take over: instead of seeking to maximize a certain objective cost function, we are likely to persist in going to a specific place just because we are used to it. Anecdotal evidence shows that this is indeed what often happens in practice! 


Our model assumes that the utility landscape is affected by past choices, with a memory kernel that decays with time. Agents can change their decision using a logit (or Metropolis) rule, parameterised by an ``intensity of choice'' $\beta$ that plays the role of the inverse temperature in statistical physics. This type of model belongs to a wide class of so-called ``reinforcement'' models, which contains Polya Urns, Reinforced Random Walks, Elephant Walks, etc. -- for a review see \cite{pemantle2007} and references therein.\footnote{For recent developments, see also \cite{Boyer_2019,Jack}.} Such models have also gained traction in the economics literature, where positive reinforcement of certain choices made by agents are shown to impact the emergence of certain macro outcomes and structures \cite{BrianArthur1987,ArthurPosFeedback}.

After properly defining our model, we provide analytical arguments to confirm the intuition that sufficiently strong memory effects, coupled with the optimization of the subjective memory-induced utility, can lead to ``self-trapping'', i.e. the agent sticks to a choice whose objective utility is not necessarily maximal, simply by force of habit. We confirm our result \textit{via} numerical simulations that explore different topologies for the space of different choices. We discover a particularly interesting dynamical transition when the memory kernel decays as the inverse of time, with rather unusual scaling and super-aging properties. We finally propose possible extensions. 


\section{A Simple model}

Consider a set of $N$ discrete choices, labeled $(x_i)_{1\leq i \leq N}$, to which we assign an utility $-$ a measure of the value an individual assigns a given choice. The perceived utility of site $x_i$ and time $t$ is postulated to be: 
\begin{equation}\label{eq:utility_form}
U(x_i,t) = U_0(x_i)\left(1+\sum_{t'=0}^t \phi(t-t') \mathbf{1}_{x(t')=x_i}\right)  ,
\end{equation}
where the first term on the right-hand side is the intrinsic, or objective utility of the choice, while the second accounts for memory effects, affecting the utility of that choice for the only reason that the individual has picked it in the past.\footnote{One may also think, in the physicist's language, of an energy landscape (akin to minus the utility) where the energy of a given site or configuration increases or decreases if the system has already visited that site.}  The  decaying memory kernel $\phi$ encodes that more recent choices have a stronger effect, and $x(t)$ denotes the choice of the individual at time $t$. Hence, past history ``chisels'' the utility landscape, in a way similar to ants leaving a pheromone trace that guide other ants along the same path, or rivers creating their own bed through erosion. Note that in most reinforcement random walk models reviewed in \cite{pemantle2007}, infinite memory span is assumed, i.e. $\phi(t)=$ constant, while we will be mostly concerned here with decaying memory kernels.

The sign of the kernel $\phi$ separates two different cases: $\phi<0$ indicates a situation where an individual grows weary of his past choices, while $\phi>0$ corresponds to the case where an individual becomes increasingly endeared with them. In agreement with intuition, the former case leads to an exploration of all the choices unless the optimal choice has an utility too far apart from the rest to be sufficiently affected by the kernel. In all that follows we focus on the more interesting case $\phi\geq 0$. The reason behind studying such utility reinforcement lies in the behavioural idea that people tend to prefer what they already know, thus paving the way for ``habit formation" as in~\cite{kirmanfishes2000}, see also \cite{abel1990asset,constantinides1990habit,carroll2000saving, fuhrer2000habit,pol1970sav} and \cite{bouchaudcrisis2013}. 
We then consider the following dynamics. An individual, standing by choice $x_i$ at time $t$, draws an alternative $x_j$ from a certain ensemble of ``nearby choices'' $\partial_i$, e.g. the set of neighbors of $i$ in a graph $\mathcal{G}$, with probability: 
\begin{eqnarray}
\mathcal{T}_{x_i\to x_j} = \frac{\mathbf{1}_{x_j\in\partial_i}}{{\mathcal N}_i}, \quad \mathcal{N}_i:={\sum_j \mathbf{1}_{x_j\in\partial_i}},
\end{eqnarray}
where $\mathcal{N}_i$ is the number of neighbours of $i$.
Restricting to nearby choices is a parsimonious way to model adaptation costs, that penalize large decision changes. However, our framework is quite versatile since the topology of the graph $\mathcal{G}$ is arbitrary, and we will consider different cases below. 

\begin{figure}[tb]
    \centering
    \includegraphics[width=0.5\textwidth]{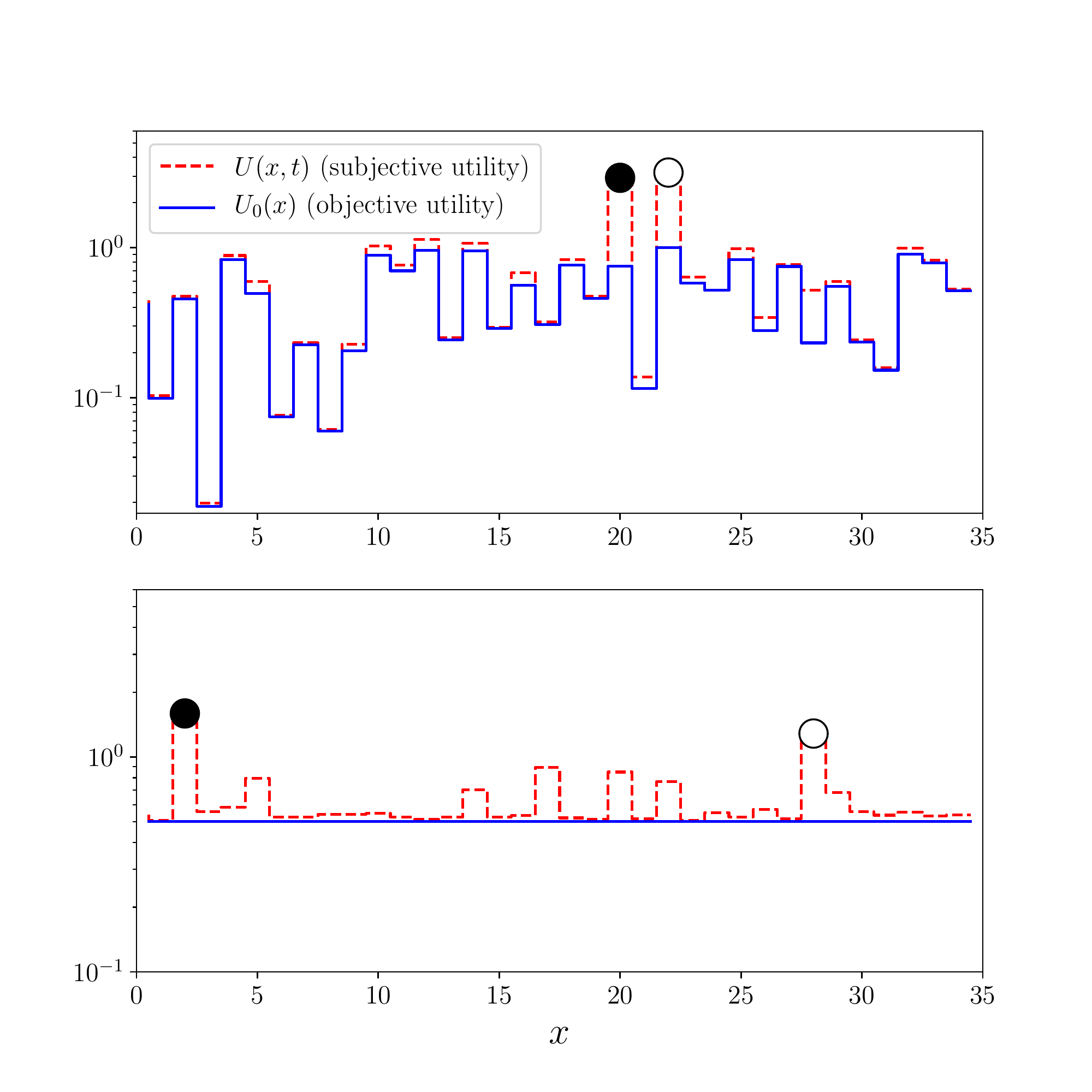}
    \caption{Schematic representation of our problem at a given time $t$. The plot on the top depicts the case of random ``objective'' utilities $U_0(x)$, while the one on the bottom shows the situation where they are uniform $U_0(x)=U_0$. In both plots, the solid black ball represents the choice made at time $t$, while the empty ball represents the choice made at time $t-1$. Both correspond to a simulation run with a power-law kernel $\phi(t)\propto (1+t)^{-\gamma}$ with $\gamma=1.5$ and $\beta=0.2$, on a fully connected graph.}
    \label{fig:illustration}
\end{figure}

The target choice $x_j$ is then adopted with the logit probability, standard in Choice Theory~\cite{anderson1992discrete}:\footnote{For non-trivial trapping to emerge, we consider graphs without singletons, that is to say that all sites have a non-empty set of neighbours that are different to itself.}
\begin{equation}\label{eq:logit}
p(x_i\to x_j) = \frac{1}{1+e^{\beta [U(x_i,t)-U(x_j,t)]}},
\end{equation}
where $\beta$ is called the ``intensity of choice'' and accounts for the degree of rationality (it is the analogue of the inverse temperature in statistical mechanics). Indeed, as long as $0< \beta<\infty$, the agent is more likely to switch whenever $U(x_j,t)>U(x_i,t)$ (optimizing behaviour), but the probability to pick a choice with a lower utility is non-zero, which encodes for bounded-rationality (or uncertainty about the true utility) in the economics literature. In the $\beta \to 0$ limit (equivalent to the infinite temperature limit in physics) the agent explores the whole space of possible choices without taking their utility into account. In the opposite limit $\beta\to\infty$ (or zero temperature) the agent has a greedy behaviour and only switches to choices with a higher utility, but this also implies that he/she may stay in a local maximum instead of taking the chance to explore all available possibilities. An illustration of these dynamics is given in Fig.~\ref{fig:illustration}.

When $\phi =0$, the dynamics is that of the Metropolis-Hastings algorithm used to sample the Boltzmann-Gibbs distribution~\cite{Metropolis1949,Hastings1970}. The stationary state of the dynamics is such that the probability to pick choice $x_i$ is proportional to $\mathcal{N}_i e^{\beta U_0(x_i)}$. This can by itself lead to interesting phenomena depending on the statistics of $U_0$. For example the study of the Random Energy Model~\cite{DerridaREM, bouchaud_mezard}, for a finite value of $N$ and for Gaussian utilities of variance $\sigma^2$, shows that for $\beta>\beta_c = {\sqrt{2\ln{N}}}/{\sigma}$, the probability measure condenses on a small number of choices, much smaller than $N$.  

\section{Non-Ergodicity \& Condensation of Choices}

Adding the kernel introduces the possibility that the agent gets stuck in a non-optimal choice exclusively through memory effects: staying a long time in a given choice self-reinforces its utility, thereby increasing the likelihood to stay there and leading to non-ergodic dynamics. To study the possible condensation or trapping induced by memory alone, we restrict ourselves to the case where $U_0(x_i)=1,\ \forall i$. The interplay between memory-induced trapping and utility heterogeneity is quite interesting in itself, but we leave it to future investigations.  

We consider an agent starting from a given choice $x_0$ at time $t=0$ and follow his/her evolution for times $t=1,\ldots,T$ with $T$ sufficiently large. We then compute the empirical state histogram $p_i=\sum_{t} \mathbf{1}_{x(t)=x_i}/T$ and define the order parameter $h$, in a similar way to the inverse participation ratio used in condensed matter physics~\cite{Wegner1980,Bell1970} or to the Herfindahl index in economics~\cite{Hall1967}, as:
\begin{equation}\label{eq:herfindahl}
h := \sum_{i=1}^N p_i^2.
\end{equation}
This parameter indicates how the agent has explored the space of possible choices: if all choices were visited with equal probability then one has immediately $p_i = 1/N$ and thus $h = \sum_i 1/N^2 = 1/N$. On the other hand if the agent was stuck in a single choice $j$, then $p_i=\delta_{i,j}$ and so $h=1$. Therefore $1/h$ gives an order of magnitude of the number of different choices picked by the agent during time $T$. In practice, we average $h$ over a large number of trajectories and starting sites $x_0$, to obtain an average parameter $\avg{h}$. For a set of simulations on a graph $\mathcal{G}$ with $N$ choices and lasting a time $T$, we therefore define the critical value  $\beta_c$, defining the crossover between $h=\mathcal{O}(1/N)$ and $h=\mathcal{O}(1)$ as the value for $\beta$ that maximizes the variance of $h$ over different trajectories. 

An important question is whether $\beta_c$ corresponds to a true transition or to a mere crossover. This depends on the $L^1$ norm of the memory kernel, $\norm{\phi} = \sum_{t=0}^\infty \phi(t)$. Suppose that this norm is finite. Then if the agent has been stuck in a given site $i$ for a time $t\gg 1$, we can approximate its utility by $U_0(1+\norm{\phi})$. The difference in utility with the neighbouring choices thus remains finite. For any finite value of $\beta$, the probability to leave that site is non-zero, and therefore the individual will eventually pick a different choice. The time for this to happen is however of the order of $\exp(\beta U_0 \norm{\phi})$. If this time is much longer than $T$, we will in fact measure $h \sim 1$, even though running the trajectory for a longer time would result in $h=\mathcal{O}(1/N)$. Hence in this case $\beta_c$ is a crossover that depends on $T$ as $\ln(T)/(U_0 \norm{\phi})$. 

A more interesting situation (at least from a theoretical point of view) is when $\norm{\phi}=\infty$. As we will show below, there exists cases where $\beta_c$ corresponds to a true phase transition and is independent of $T$ (when $T$ is large). 

\section{Mean Field Approximation}

In order to draw further analytical features, we start by looking at a mean-field approximation. This means that we take the graph $\mathcal{G}$ to be fully connected with $\mathcal{T}_{x_i\to x_j} = 1/(N-1)$ and in the limit $N\to\infty$. 

We now formalize the argument previously sketched. If the individual started first at a given choice corresponding to node $i$, then the probability $P_>(\tau)$ that he/she remains there up to a time $\tau$ is given by the product over $ t \in\llbracket 0, \tau-1\rrbracket$ of the probabilities not to leave the site between times $t$ and $t+1$, $p_{\text{stay}}(t)$. Now, $ p_{\text{stay}}(t) =1 - p_{\text{leave}}(t)$ with: 
\begin{equation}\label{eq:stay}
 p_{\text{leave}}(t) = \sum_{j\in \partial_i}\mathcal{T}_{x_i\to x_j} \frac{1}{1+e^{\beta[U(x_i,t)-U(x_j,t)]}}  .
\end{equation}
For the fully connected graph, this expression simplifies to\footnote{In the general case in which the agent started in a ``site'' different from $i$ and then got stuck in $i$, one wants to replace the right-hand side by an average of the logit rule over the utility gaps $U(x_i,t)-U(x_j,t)$. However, as $N\to\infty$ it is very unlikely that a site that was previously picked is chosen again. We can thus safely replace the average gap by the gap with the base level $U_0=1$.} $p_{\text{stay}}(t) = \big[1+e^{-\beta \Phi(t)}\big]^{-1}$, with $\Phi(t) =U_0 \sum_{0}^t \phi(s)$. It follows that:
\begin{equation}\label{eq:p_tau}
\begin{split}
\hspace{-2mm} P_>(\tau) &= \prod_{t=0}^{\tau}\left[1+e^{-\beta \Phi(t)}\right]^{-1} 
\approx e^{-I(\tau)},
\end{split}
\end{equation}
with $ 
I(\tau) := \int_{0}^{\tau} \mathrm{d}t\,\ln\left[1+e^{-\beta \Phi(t)}\right]$,
where we have replaced discrete sums by integrals.
Equation~\eqref{eq:p_tau} determines the distribution of the ``trapping'' time $\tau$ that the agent spends stuck on a certain choice. Its nature will entirely depend on the behaviour of the integral $I(\tau)$  
when $\tau \to \infty$. 
\subsection*{Short Term Memory}
Consider first the case where $\lim_{t \to \infty} \Phi(t) = \norm{\phi}<+\infty$. Then $I(\tau) \approx \lambda \tau$ for $\tau \to \infty$, with 
\begin{equation}\label{eq_lambda}
\lambda:= \ln\left[1+e^{-\beta \norm{\phi}}\right].
\end{equation}
This means that the trapping time distribution decays exponentially fast for large $\tau$, with an average trapping time $\langle \tau \rangle$ approximately given by $1/\lambda$. For sufficiently small $\lambda$, we recover the qualitative criterion of the previous section by setting $T \lambda \sim 1$. But the dynamics remains ergodic when $T \to \infty$.

\subsection*{Long Term Memory}
Suppose now that $\phi(t)$ decays sufficiently slowly for large $t$ for $\norm{\phi}$ to diverge. For definiteness, we will focus on power-law kernels:
    \begin{equation}\label{eq_phi_gamma}
          \phi(t)= \frac{C}{(1+t)^\gamma}.
    \end{equation}
    When $\gamma > 1$, $\norm{\phi}$ is finite and we are back to the previous case. Hence we restrict to $\gamma \leq 1$.
    When $\gamma < 1$, one finds that $\Phi(t) \propto t^{1-\gamma}$ for large $t$. Hence $I(\tau)$ converges to a finite limit $I_\infty$ for large $\tau$. This means that there is a finite probability $P_\infty=e^{-I_\infty}$ that the choice is made {\it forever}.
    When $\gamma=1$, $\Phi(t) \approx CU_0 \ln t$ for large $t$. This leads to three further sub-cases:
    \begin{enumerate}
        \item When $\beta C U_0 > 1$, $I(\tau)$ again converges to a finite limit when $\tau \to \infty$, i.e. decisions self-trap forever.
        \item At the transition point, defined as $\beta_c^\star=(CU_0)^{-1}$, one finds that $P_>(\tau)$ decays as $\tau^{-1}$, i.e. the trapping time distribution is a Zipf law, $\tau^{-2}$. This is the marginal case that appears in several models of aging in the literature \cite{Bertin_2002,amir2012}. For a finite observation time $T$, the average trapping time grows like $\ln T$.
        \item When $\beta C U_0 < 1$, $I(\tau)$ behaves for large $\tau$ as $\exp(-\tau^b/b)$, where $b=1 - \beta C U_0 > 0$. The average trapping time $\langle \tau \rangle$ is thus finite. A careful analysis shows that $\langle \tau \rangle$ diverges as $b^{-1}$ when $b \to 0$, but higher moments $\langle \tau^k \rangle$ with $k > 1$ diverge much faster, as $\exp((k-1)/b)$, i.e. according to the so-called Vogel-Tamman-Fulcher law, see e.g. \cite{Debenedetti2001}.
    \end{enumerate}

Let us summarize the above results. When the kernel $\phi$ decays fast enough, there is a crossover regime in $\beta$ between free exploration of the space of choice and trapping. The crossover value of $\beta$ depends on the observation time $T$ and is given, using Eq. \eqref{eq_lambda} for $T$ large, by
\begin{equation}
    \label{eq:transition}
    \beta_c = \frac{\ln T}{\norm{\phi}}.
\end{equation}
When memory is long ranged, and described by a power-law kernel with decay exponent $\gamma$, there exists a genuine transition when $\gamma=1$ between a free exploration regime and a (non-ergodic) trapped regime at a $T$ independent value of $\beta$ that we shall henceforth call $\beta_c^\star$. When $\beta > \beta_c^\star$ or $\gamma < 1$, there is a non-zero probability to get trapped in the same decision forever. The characteristic time for changing decision is of the same order of magnitude as $T$ itself, a phenomenon called ``aging'', see e.g. \cite{Bouchaud1992, amir2012}, on which we will comment further below, see Fig.~\ref{fig:aging}. Note finally that for $\gamma=1$, our mean-field analysis predicts that while the average trapping time diverges as $(\beta_c^\star - \beta)^{-1}$, all higher moments diverge much faster, as $\sim \exp(A/(\beta_c^\star - \beta))$.

\section{Numerical results}

We have conducted simulations using a long-range memory kernel given by Eq. \eqref{eq_phi_gamma} with $\gamma\in[1;\infty[$. For numerical convenience, we represent $\phi(t)$ as a superposition of exponentials as done in \cite{bochud2006optimal}. We have considered a variety of different graph topologies $\mathcal{G}$: fully connected graphs, one dimensional chains, and finally Watts-Strogatz small world networks. Without loss of generality, we set $U_0=1$ as this simply corresponds to a rescaling of $\beta$.

\begin{figure}[t!]
    \centering
    \includegraphics[width=1\columnwidth]{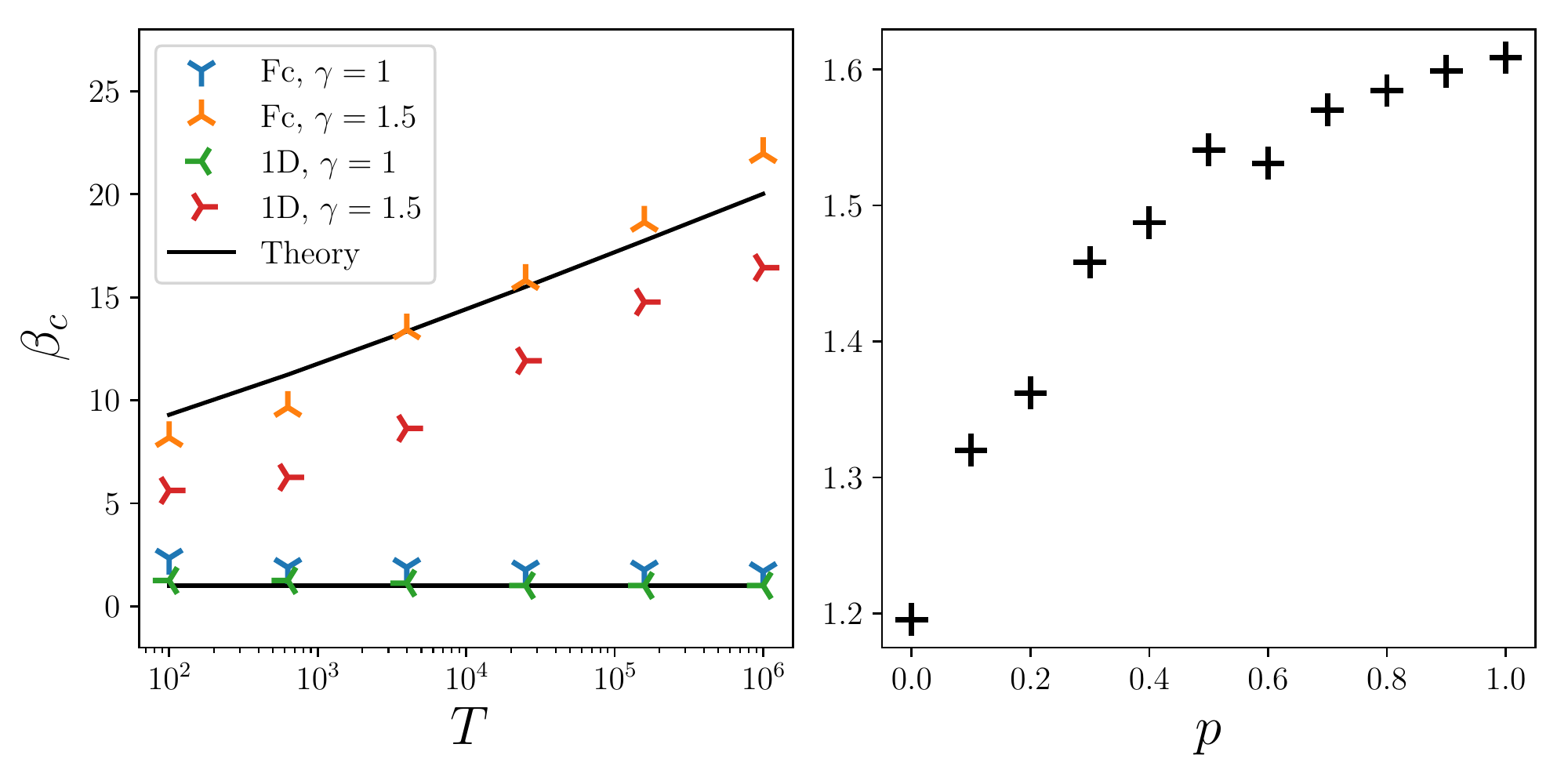}
    \caption{Left: critical $\beta_c$ as a function of $\ln T$ for $\gamma=1$ and $1.5$, $N=10^5$ and different topologies. (Fc stands for fully connected, while 1D is the one-dimensional chain). Black lines correspond to the prediction of mean-field theory. Right: dependence of the critical intensity of choice $\beta_c^\star$ on the parameter $p$ of Watts-Strogatz networks, for $T=5\cdot 10^3$ and $N=2\cdot 10^3$.}
    \label{fig:betac}
\end{figure}

Figure~\ref{fig:betac} (Left) shows the value of $\beta_c$, determined as the maximum of the variance of $h$, as a function of $T$ for two different topologies (one dimensional and fully connected) and two different values of $\gamma\in\{1,1.5\}$. Our results show excellent qualitative agreement with the theoretical prediction for the two topologies, in particular  Eq.~\eqref{eq:transition} in the $\gamma>1$ case, although there is an overall factor needed to account for the one-dimensional data.

One can actually interpolate between the two situations by considering  Watts-Strogatz small-world networks~\cite{Watts1998}, with a rewiring parameter $p$ such that $p=0$ corresponds to one-dimension chains and $p=1$ to  the fully connected graph. Figure~\ref{fig:betac} (Right) shows the value of $\beta_c^\star$ as a function of the rewiring parameter $p$ of interpolating between between one-dimensional chains for $p=0$ and the fully connected graph for $p=1$.  The parameter $p$ therefore allows to interpolate between a situation where one may only do local jumps to a situation where one can go anywhere. As expected, $\beta_c^\star$ increases with $p$, as it is easier to get trapped in less connected graphs, where the same choice is revisited more often.  

\begin{figure}[b!]
    \centering
    \includegraphics[width=1\columnwidth]{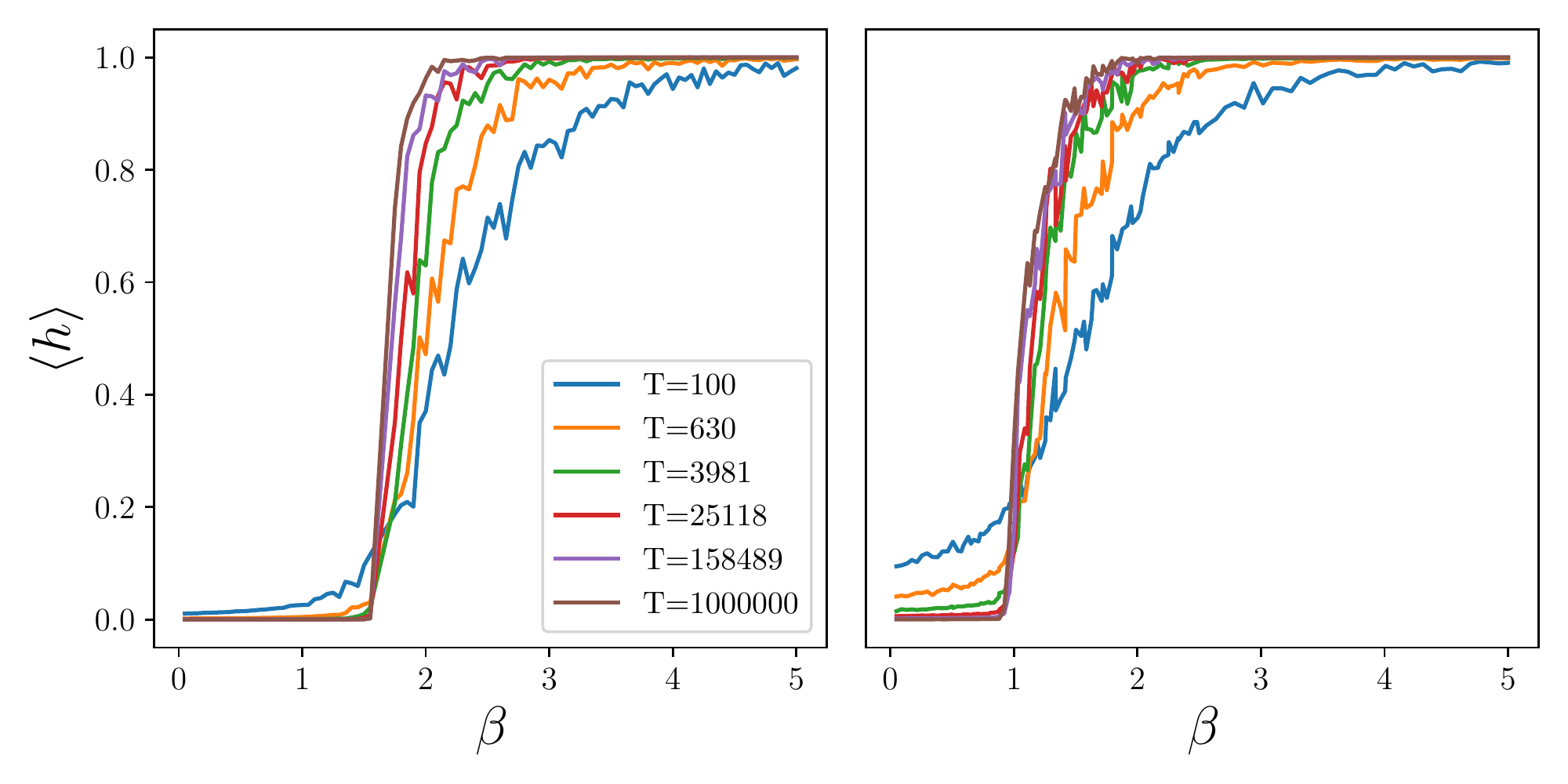}
    \caption{Order parameter $h$ as a function $\beta$ for $N=10^5$, $\gamma =1$ and different values of $T$. Left: Fully connected graph, for which a step function is approached as $T^{-1/3}$. Right: One-dimensional chain geometry, for which we cannot exclude that $h$ remains a continuous function of $\beta$ when $T \to \infty$.}
    \label{fig:super}
\end{figure}

 We now study more carefully the behaviour of the order parameter $h$ close to the transition point $\beta_c^\star$, when $\gamma=1$, both for one dimensional chains and for the fully connected graph. We choose $N=10^5$ henceforth, such that finite size effects are negligible in the range of $T$ that we explore. Figure~\ref{fig:super} suggests that as $T \to \infty$,  $\langle h \rangle(\beta)$ appears to slowly converge to a step function that is zero for $\beta < \beta_c^\star$ and unity when $\beta > \beta_c^\star$, at least in the fully connected case where the speed of convergence is found to be $\sim T^{-1/3}$. In the one dimensional case, one cannot exclude with the available data that this limiting function remains continuous when $T \to \infty$.  

\section{Aging}

\begin{figure}[t!]
    \centering
    \includegraphics[width=1.03\columnwidth]{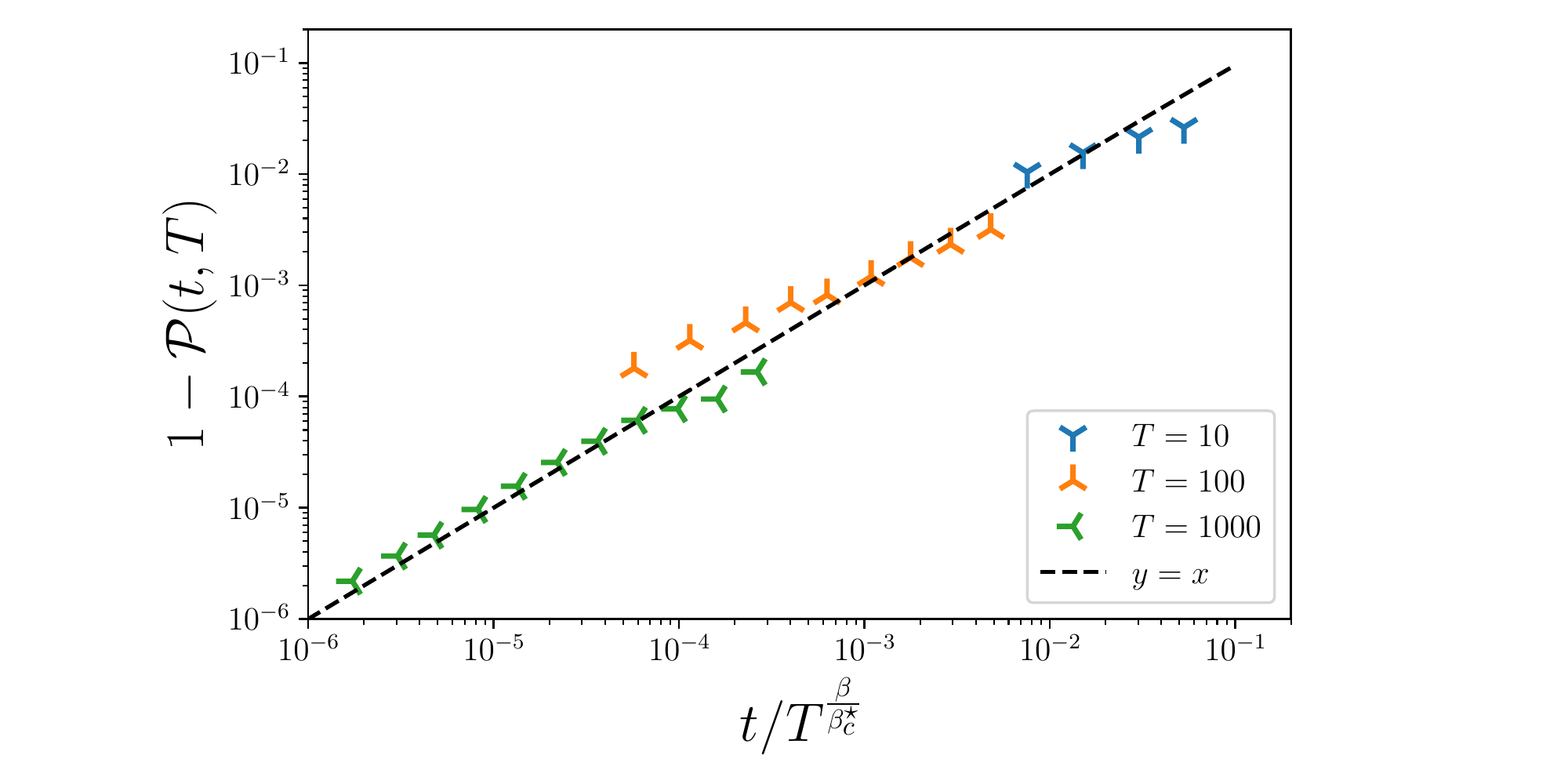}
    \caption{Aging function $1-\mathcal{P}(t,T)$ as a function of $t/T^{\beta/\beta_c^\star}$ in a fully connected graph with $N=10000$, with $\beta_c^\star \approx 1.5$ and $\beta =3.5$.  Eq.~\eqref{eq:aging_analytical} predicts that $\mathcal{P}(t,T) \approx 1 - t/T^{\beta/\beta_c^\star}$ when $t \ll T$ (dashed line).}
    \label{fig:aging}
\end{figure}

Finally, let us be a little more specific about the meaning of self-trapping for finite $T$ when $\beta > \beta_c^\star$. The correct statement is that the system {\it ages}, in the following sense \cite{Bouchaud1992, Monthus_1996}: assume that the agent's choice at time $T$ is a certain $x_i$ and ask: What is the probability $\mathcal{P}(t,T)$ that the agent has never changed his/her mind between $T$ and a later time $T+t$? In the free exploration phase $\beta < \beta_c^\star$, this probability is, for large $T$, independent of $T$: the process is time-translation invariant. In the trapped phase $\beta > \beta_c^\star$, $\mathcal{P}(t,T)$ can be estimated by appropriately generalizing Eq. \eqref{eq:p_tau}. The result takes the following aging form (see Fig~\ref{fig:aging}):
\begin{equation}\label{eq:aging_analytical}
\mathcal{P}(t,T) \approx \exp\left(\frac{1}{a(T+t)^{a}}-\frac{1}{aT^{a}}\right), \quad a =  \frac{\beta}{\beta_c^\star} - 1 > 0.
\end{equation}
Note that in the regime $t \ll T$, $\mathcal{P}(t,T)$ is a function of $t/T^{1+a}$ (precisely $1- \mathcal{P}(t,T)\approx t/T^{1+a}$, see dashed line in Fig.~\ref{fig:aging}), a regime called {\it super-aging} \cite{bouchaud2000aging} since the effective time for changing one's mind grows as $T^{\beta/\beta_c^\star}$, i.e. faster than the age $T$ itself. 
This is quite interesting since we are not aware of simple models leading to such a super-linear aging behaviour. Hence, memory effects of the type discussed here might very well be an interesting lead to interpret experiments that show such a super-aging behaviour, such as those reported in \cite{cipelletti2000}. 

\begin{table*}
\centering

\renewcommand{\arraystretch}{1.3}
\begin{tabular}{|l||c|c|c|}
\hline 
\begin{tabular}[c]{@{}l@{}} Asymptotics of  $\sum_{t=0}^{T}\phi(t)$\end{tabular} & $\abs{\phi}< \infty$ & $\ln(T)$ & $T^{1-\gamma},\, \gamma<1$ \\[2pt] \hline

 \begin{tabular}[c]{@{}c@{}}Asymptotics of  $P_>(\tau)$ \end{tabular}&
 \begin{tabular}[c]{@{}l@{}} \,$e^{-\lambda \tau}$ with 
 $\lambda = \ln\left[1+ e^{-\beta \abs{\phi}}\right]$
 \end{tabular} &
 
 \begin{tabular}[c]{@{}l@{}} \tabitem$\beta>\beta_{c}^{\star}$  (trapped regime):  $P_{\infty}>0$ \\
 \tabitem $\beta=\beta_{c}^{\star}$  (critical regime): $\tau^{-1}$ \\
 \tabitem  $\beta <\beta_c^{\star} $  (ergodic regime):
 $\exp(-\tau^{b}/b),$  \\ \hspace{10pt} with $b = 1-\beta/\beta_c^{\star}$ 
 \end{tabular} & 
 
\begin{tabular}[c]{@{}l@{}}$P_\infty>0$ \\ (trapped) \end{tabular}\\
\hline
Aging $\mathcal{P}(t,T)$ & \begin{tabular}[c]{@{}l@{}} Trapping \& aging for  $T \ll e^{\beta \vert \phi\vert}$ \end{tabular}   & 
 \begin{tabular}[c]{@{}l@{}} \tabitem$\beta>\beta_{c}^{\star}$ super-aging $t \sim T^{\beta/\beta_c^\star}$\\
 \tabitem $\beta=\beta_{c}^{\star}$ normal-aging $t \sim T$\\
 \tabitem  $\beta <\beta_c^{\star} $ equivalent to
  $\abs{\phi}<\infty$ \hspace{1.1cm}
 \end{tabular} & \begin{tabular}[c]{@{}l@{}} Quasi-frozen \\ relaxation \end{tabular}
 \\ \hline
\end{tabular}
\caption{Summary of the different dynamical regimes.}
\label{table:summary}
\end{table*}

Right at the transition point $\beta = \beta_c^\star$, one finds simple aging, i.e. a scaling function of $t/T$: 
\begin{equation}
\mathcal{P}(t,T) \approx \frac{1}{1+\frac{t}{T}}, \qquad  {\beta}={\beta_c^\star}.
\end{equation}
When the kernel has a finite norm and leads to a crossover rather than a true transition, aging will take place whenever $T \ll e^{\beta \norm{\phi}}$ but revert to a normal time translation dynamics when $T \gg e^{\beta \norm{\phi}}$ (see \cite{Monthus_1996} for a similar situation). When $\gamma < 1$, on the contrary, relaxation is quasi-frozen for large $T$, in the sense that $\mathcal{P}(t,T) \approx 1 - t \exp(-\beta U_0 T^{1-\gamma})$ when $t \ll T^\gamma$.

\section{Conclusion}   

Although quite simple, our model shows that non-trivial choice distortion effects can emerge through memory or self-reinforcing mechanisms. 
Our main result is that the addition of memory effects can hinder the full exploration of all choices by the agent, and it may even cause him/her to leave a substantial number of possible options totally unexplored. The emergence of aging properties also shows that including memory effects in agents' preferences can lead to non-ergodic dynamics, when ergodicity is a crucial assumption to many models in economics. Table~\ref{table:summary} summarises our results.

Several extensions can be thought of, and would be a sensible way to incorporate more realism into the model. As we mentioned, we have explored here the case where the objective utility landscape $U_0(x)$ is totally flat, in a way to highlight the effects induced by memory alone. Reintroducing some heterogeneities in $U_0(x)$ would allow one to study the competition between ``landscape trapping'' and ``memory trapping'', with possibly interesting transitions between the two. Another direction is to introduce many agents with interactions between them, meaning that the subjective utility may also depend on what others are doing. Here again, one expects some interesting competition between herding induced condensation of choices and memory effects.  In particular, the combined effect of imitation of the past and imitation of peers may generate \textit{collective self-fulfilling prophecies}. 

 Another direction one could explore is when the graph $\mathcal{G}$ defining the topology of the space of choices is itself time dependent -- see \cite{Cotar2009} for a step in this direction. For example, the neighbourhood of each choice could be itself affected by past choices, or some new choices, not present initially, could present themselves later in time (for example, the opening of a new restaurant). 

In all these cases, the basic question is whether memory effects, habit formation, or herding completely distort the choices dictated by their objective utilities or not. Such distortions may have very significant economic consequences at the macro level. 

From a purely theoretical point of view, revisiting reinforcement models considered in the literature \cite{pemantle2007} with a power-law decaying memory kernel could lead to new interesting transitions of the type discussed above. In particular, the super-aging behaviour reported in the trapped phase might have applications much beyond the present setting. 

\vskip 0.2cm
\emph{Acknowledgements.} We warmly thank Pierre Lecointre, who contributed to the early stages of this project, as well as Jean-Pierre Nadal and Alan Kirman for their suggestions. This research was conducted within the \textit{Econophysics \& Complex Systems} Research Chair under the aegis of the \textit{Fondation du Risque}, a joint initiative by the \textit{Fondation de l'Ecole polytechnique, l'Ecole polytechnique} and Capital Fund Management.

\bibliography{biblio}   
    
\end{document}